\documentclass[10pt,conference]{IEEEtran}
\IEEEoverridecommandlockouts
\usepackage[labelsep=period]{caption}   
\usepackage{subcaption}
\usepackage{booktabs}
\usepackage{listings}
\usepackage{enumitem}
 \usepackage[most]{tcolorbox}
\usepackage{multirow}
\usepackage{graphicx}
\usepackage[normalem]{ulem}
\useunder{\uline}{\ul}{}
\usepackage{graphicx}
\usepackage{pifont}
\graphicspath{ {./fig/} }
\usepackage{url}
\usepackage{amsfonts}
\usepackage{hyperref}
\usepackage{xspace}
\usepackage{amsmath,amsfonts}
\usepackage{enumitem}
\usepackage{amssymb}
\usepackage{multirow}
\usepackage[normalem]{ulem}
\useunder{\uline}{\ul}{}
\usepackage{graphicx}
\usepackage{pifont}
\graphicspath{ {./fig/} }
\usepackage{url}
\usepackage{xspace}
\hypersetup{
	colorlinks   = true,
	urlcolor     = black, %Colour for external hyperlinks
	linkcolor    = black, %Colour of internal links
	citecolor   = black %Colour of citations
}

\AtBeginDocument{%
  \providecommand\BibTeX{{%
    \normalfont B\kern-0.5em{\scshape i\kern-0.25em b}\kern-0.8em\TeX}}}

\newcommand{\tool}{ContraBERT\space}

\begin{document}
\title{ContraBERT: Enhancing Code Pre-trained Models via Contrastive Learning}

\author{
  \IEEEauthorblockN{%
    Shangqing Liu\textsuperscript{1},\,
    Bozhi Wu\textsuperscript{1},\,
    Xiaofei Xie\textsuperscript{2}\IEEEauthorrefmark{2}\thanks{\IEEEauthorrefmark{2} Corresponding author.},\,
    Guozhu Meng \textsuperscript{3},\,
    and Yang Liu\textsuperscript{1}
  }
  \IEEEauthorblockA{%
    \textsuperscript{1}Nanyang Technological University, Singapore\\
    \textsuperscript{2}Singapore Management University, Singapore \\
    \textsuperscript{3}SKLOIS, Institute of Information Engineering, Chinese Academy of Sciences, China\\
    \{shangqin001,bozhi001\}@e.ntu.edu.sg, xiaofei.xfxie@gmail.com, mengguozhu@iie.ac.cn, yangliu@ntu.edu.sg
  }
  }

\maketitle
\begin{abstract}
Large-scale pre-trained models such as CodeBERT, GraphCodeBERT have earned widespread attention from both academia and industry. 
Attributed to the superior ability in code representation, they have been further applied in multiple downstream tasks such as clone detection, code search and code translation. However, it is also observed that these state-of-the-art pre-trained models are susceptible to adversarial attacks. The performance of these pre-trained models drops significantly with simple perturbations such as renaming variable names. This weakness may be inherited by their downstream models and thereby amplified at an unprecedented scale. To this end, we propose an approach namely ContraBERT that aims to improve the robustness of pre-trained models via contrastive learning. Specifically, we design nine kinds of simple and complex data augmentation operators on the programming language (PL) and natural language (NL) data to construct different variants. Furthermore, we continue to train the existing pre-trained models by masked language modeling (MLM) and contrastive pre-training task on the original samples with their augmented variants to enhance the robustness of the model. The extensive experiments demonstrate that ContraBERT can effectively improve the robustness of the existing pre-trained models. Further study also confirms that these robustness-enhanced models provide improvements as compared to original models over four popular downstream tasks.
\end{abstract}


\section{Introduction}
\label{sec-intro}
It has already been confirmed that the ``big code'' era~\cite{allamanis2018survey} is coming due to the ubiquitousness of software in modern society and the accelerated iteration of the software development cycle (design, implementation and maintenance). According to a GitHub official report~\cite{GitHub} in 2018, GitHub has already reached 100 million hosted repositories. The Evans Data Corporation~\cite{lu2021codexglue} also estimated that there are 23.9 million professional developers in 2019 and that number is expected to reach 28.7 million in 2024. As a result, the availability of code-related data is massive (e.g., billions of code, millions of changed code, bug fixes and code documentation), which yields a hot topic in both academia and industry. That is how to adopt the data-driven approach (e.g., deep learning) to solve conventional software engineering (SE) problems.

Deep learning has been widely applied to diverse SE tasks (AI4SE) such as software vulnerability detection~\cite{allamanis2017learning, zhou2019devign, allamanis2021self}, source code summarization~\cite{liu2020retrieval, alon2018code2seq, alon2019code2vec}, deep code search~\cite{liu2021graphsearchnet, gu2018deep} and source code completion~\cite{svyatkovskiy2021fast, alon2020structural, liu2020multi}. Besides, the early works~\cite{iyer2016summarizing, husain2019codesearchnet, russell2018automated, barone2017parallel, zhou2021spi} directly utilized vanilla deep learning techniques such as Long-Short Memory Networks (LSTMs)~\cite{hochreiter1997long} and Convolutional Neural Networks (CNNs)~\cite{krizhevsky2012imagenet} for different tasks. Later works~\cite{zhou2019devign, liu2020retrieval, liu2021graphsearchnet, liu2020atom, wu2022enhancing, li2022transrepair, allamanis2017learning, alon2018code2seq} customized different network architectures to satisfy the characteristics of the specific task for achieving the best performance. For example, since complicated data dependencies and control dependencies are easier to trigger software vulnerabilities, Devign~\cite{zhou2019devign} incorporated different kinds of program structure information with Code Property Graph~\cite{yamaguchi2014modeling} to Graph Neural Networks~\cite{li2015gated} for vulnerability detection. Considering code duplication~\cite{allamanis2019adverse} is common in the ``big code'' era, Liu et al.~\cite{liu2020retrieval} combined the retrieved code-summary pair to generate high-quality summaries. Although these customized networks have achieved significant improvements on specific tasks, the generalization performance is still low. To address this limitation, some researchers propose to utilize unsupervised techniques with the massive amount of data to pre-train a general model~\cite{kanade2019pre, feng2020codebert, guo2020graphcodebert, lu2021codexglue, svyatkovskiy2020intellicode,buratti2020exploring, karampatsis2020scelmo, wang2021codet5, ahmad2021unified} and then fine-tune it for different downstream tasks. For example, CuBERT~\cite{kanade2020learning} pre-trained BERT~\cite{devlin2018bert} on a large collected Python corpus (7.4M files) and then fine-tuned it on different tasks such as variable-misuse identification and wrong binary operator identification. CodeBERT~\cite{feng2020codebert} pre-trained RoBERTa~\cite{liu2019roberta} for programming languages (PL) with their natural language (NL) comments on the open-source six programming languages~\cite{husain2019codesearchnet} and evaluated it on code search and source code summarization. GraphCodeBERT~\cite{guo2020graphcodebert} further incorporated data flow information to encode the relation of variables in a program for pre-training and demonstrated its effectiveness on four downstream tasks. 
% In addition, GraphCodeBERT is considered as the first work to leverage the semantic structure of the code for the pre-training. 
% Furthermore, they also implemented CodeGPT based on GPT-2~\cite{radford2019language} for code completion and text-to-code generation tasks and released a machine learning benchmark (CodeXGLUE~\cite{lu2021codexglue}) with 14 datasets from 10 diversified tasks for model evaluation and comparison.

\begin{figure}
     \centering
    \includegraphics[width=0.35\textwidth]{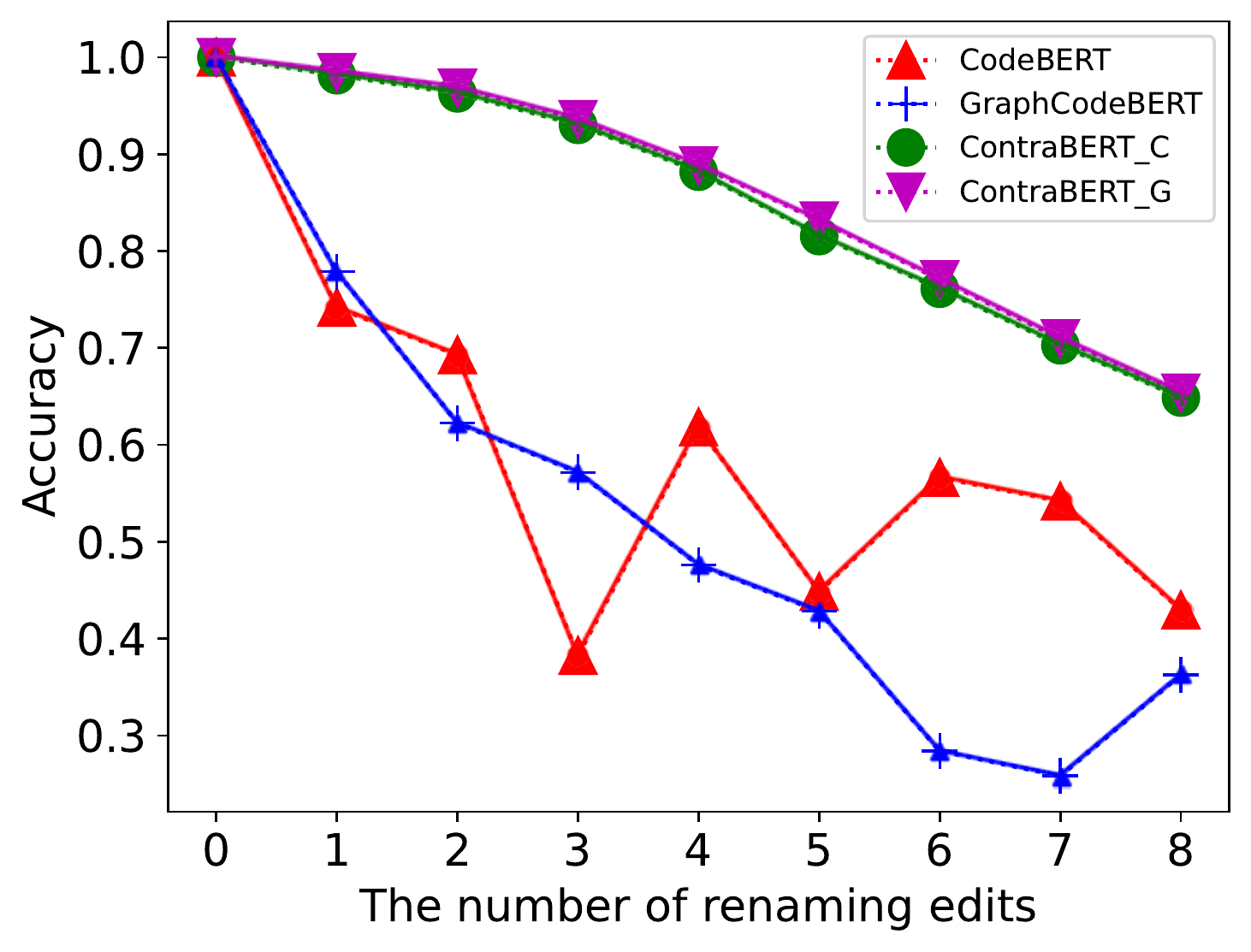}
     \caption{Adversarial attacks on clone detection(POJ-104).}
    \label{fig:robust}
\end{figure}

The aforementioned pre-trained models have a profound impact on the AI4SE community and have achieved promising results on various tasks. With the widespread use of pre-trained models, an important question is whether these models are robust to represent code semantics.
% Hence, we may ask a question: ``Is it enough to utilize these existing pre-trained techniques to represent code?'' 
% The answer is no and by our preliminary experiments, 
Our preliminary study has demonstrated that state-of-the-art pre-trained models are not robust to a simple label-preserving program mutation such as variable renaming. 
Specifically, we utilize the test data of clone detection (POJ-104)~\cite{mou2016convolutional} (a task to detect whether two functions are semantic equivalence with different implementations) provided by CodeXGLUE~\cite{lu2021codexglue} and select those samples that are predicted correctly by the pre-trained CodeBERT~\cite{feng2020codebert} and GraphCodeBERT~\cite{liu2019roberta}. 
Then we randomly rename variables within these programs from 1 to 8 edits. For example, %8 edits represent we randomly select 8 different identifiers to rename with the new generated names for all the occurrences in the function.
8 edits mean that we randomly select 8 different variables in a function and rename them for all occurrences with the newly generated names. If one function has less than 8 variables, we will rename %to
the maximum number of variables. We then utilize these newly generated mutated variants to evaluate the model prediction accuracy based on the cosine similarity of the embedded vectors of these programs. Surprisingly, we find that either CodeBERT or GraphCodeBERT suffers greatly from renaming operation and the accuracy reduces to around 0.4 when renaming edits reach to 8 (see Fig.~\ref{fig:robust}). It confirms that pre-trained models are not robust to adversarial examples. However, it is challenging to improve the robustness of pre-trained models. Although the latest work by Yang et al.~\cite{yang2022natural} proposed some attack strategies to make CodeBERT and GraphCodeBERT have poor performance on adversarial samples. They further combined adversarial samples with original samples to fine-tune pre-trained models without any changes to the model architecture to improve prediction robustness on downstream tasks. However, a newly designed model that inherently solves the weakness of robustness is not involved in their paper. 
% Recently, contrastive learning, which aims at minimizing the distance between the vectors of similar samples while maximizing the distance between the dissimilar examples, has achieved significant improvements for improving model robustness in computer vision (CV)~\cite{fan2021does, kim2020adversarial}. 
% Furthermore, it has also be confirmed can improve the model robustness~\cite{kim2020adversarial} in image recognition. 
% For example, RoCL~\cite{kim2020adversarial} has proved that contrastive learning can effectively improve the robustness of the pre-trained models from adversarial attacks in image recognition.
%By collecting or constructing similar and dissimilar samples, the objective of contrastive learning is to minimize the distance between the representations of similar samples while maximizing the distance between the dissimilar examples to obtain better clusters.
%Hence, a direct idea is how to incorporate contrastive learning based on CodeBERT to improve the performance? And how to collect or construct semantic-equivalent or semantic-close data for contrastive learning in code scenario are still an open-question. 
% A direct idea is whether and how we can adopt the idea of contrastive learning in improving robustness of pre-trained models. Furthermore, whether the improvement of the robustness of the pre-trained model helps the model learn program semantics better, thereby improving the performance on downstream tasks.

%into current existing code pre-trained models such as CodeBERT and GraphCodeBERT to improve the robustness. However, the combination way and how to collect similar and dissimilar data for contrastive learning are still not well studied.

% To use the contrastive learning in code scenario, the key challenge is how towhich 
%for contrastive learning in code scenario is still an open challenge.

In this paper, we propose ContraBERT, an unsupervised contrastive learning-based framework to enhance the robustness of existing pre-trained models in code scenarios. Compared with Yang et al.~\cite{yang2022natural}, we design a new pre-trained model that takes masked language modeling (MLM) and contrastive pre-training task as the pre-training tasks to improve model robustness. To design a contrastive pre-training task to help the model group similar samples while pushing away the dissimilar samples, we define nine kinds of simple or complex data augmentation operators that transform the original program and natural language sequence into different variants. Given an existing pre-trained model such as CodeBERT or GraphCodeBERT, we take the original sample as well as its augmented variants as the input to train the model with MLM and contrastive pre-training task, where MLM is utilized to help the model learn better token representations and contrastive pre-training task is utilized to help the model group the similar vector representations to enhance model robustness. As shown in Fig.~\ref{fig:robust}, ContraBERT\_C and ContraBERT\_G denote the models are pre-trained from CodeBERT and GraphCodeBERT with our approach respectively, we observe that with the increasing number of edits, although the performance continues to drop, the curve for ContraBERT is much smoother. The prediction accuracy of ContraBERT\_C and ContraBERT\_G outperform CodeBERT and GraphCodeBERT significantly, indicating that ContraBERT\_C and ContraBERT\_G are more robust than the original models. We further perform an ablation study to confirm each type of defined PL-NL augmentation operator is effective to improve the model robustness. Finally, we conduct broad research on four downstream tasks (i.e., clone detection, defect-detection, code-to-code-trans and code search) to illustrate that these robustness-enhanced models provide significant improvements as compared to the original models. In summary, our main contributions are as follows:

\begin{itemize}[leftmargin=*]
    % \item We empirically demonstrate that current state-of-the-art pre-trained models in the code scenario such as CodeBERT, GraphCodeBERT are not robust and simple mutations reduce the performance significantly. 
    \item We present a framework ContraBERT that enhances the robustness of existing pre-trained models in the code scenario by the pre-training tasks of masked language modeling and contrastive learning on original samples as well as the augmented variants.
    \item We design nine kinds of simple or complex data augmentation operators on the programming language (PL) and natural language sequence (NL). Each operator confirms its effectiveness to improve the model's robustness.
    % \item The experiments from the open-source benchmark CodeXGLUE have confirmed that the robustness of the pre-trained models can be improved by ContraBERT. 
    \item The broad research on four downstream tasks demonstrates that the robustness-enhanced models provide improvements as compared to the original models. Our code and model are released on~\cite{website} for reproduction.
    % i.e., clone detection, defect-detection, code-to-code-trans, code-refinement and code search as compared to the initialized model.
    % We achieve new state-of-the-art results on multiple tasks.
\end{itemize}

% \noindent \textit{Soundness, Significance and Novelty.}
% ContraBERT begins at the finding that the well-known pre-trained CodeBERT in DL4SE is sensitive to adversarial attacks and even simple mutations will degrade the performance significantly. Our main novelty is to reshape the learning space by contrastive learning, which is to group the similar data and push apart the dissimilar data generated from a set of PL-NL transformations. The extensive experiments on the open-source benchmark CodeXGLUE demonstrate that the robustness and the performance have been enhanced and improved by ContraBERT. We obtain new state-of-the-art scores in multiple tasks. To the best of our knowledge, this paper is the first to explore the robustness of pre-trained model in the code scenario, and our research provides the first step.

\noindent \textbf{Organization:}
The remainder of this paper is organized as follows: Section~\ref{sec-bg} describes the background of the original models that ContraBERT will use. We elaborate our approach in Section~\ref{sec-app}. Section~\ref{sec-experimentalsetup} and Section~\ref{sec-evaluation} present the experimental setup and experimental results. In Section~\ref{sec-discussion}, we give some discussions about our work. After a brief review of related work in Section~\ref{sec-related}, we conclude this paper in Section~\ref{sec-con}.

\section{Background}
\label{sec-bg}
In this section, we briefly introduce CodeBERT and GraphCodeBERT which will be adopted as our original pre-trained models for ContraBERT.

\subsection{CodeBERT}
CodeBERT~\cite{feng2020codebert} is pre-trained on an open-source benchmark CodeSearchNet~\cite{husain2019codesearchnet}, which includes 2.1M bimodal NL-PL (comment-function) pairs and 6.4M unimodal functions without comments across six programming languages. The model architecture is the same with RoBERTa~\cite{liu2019roberta}, which utilizes multi-layer bidirectional Transformer~\cite{vaswani2017attention} for unsupervised learning. Specifically, CodeBERT consists of 12 identical layers, 12 heads and the dimension size for each layer is 768. In total, the number of model parameters reaches 125M. Two different pre-training objectives are used, the first one is masked language modeling (MLM), which is trained on bimodal data. MLM objective targets predicting the original tokens that are masked out in NL-PL pairs. To fully utilize unimodal data, CodeBERT further uses Replaced Token Detection (RTD) objective on both bimodal and unimodal samples. RTD objective is designed to determine whether a word is original or not. At the fine-tuning phase, two downstream tasks (i.e., code search and source code documentation generation) are used for evaluation. The experimental results demonstrate that CodeBERT outperforms supervised approaches on both tasks. 

\subsection{GraphCodeBERT}
GraphCodeBERT~\cite{guo2020graphcodebert} is a pre-trained model for code, which considers structures in code. Specifically, it incorporates the data flow of code to encode the relations of ``where the value comes from'' between variables in the pre-training stage. In addition to the pre-training task of masked language modeling (MLM), GraphCodeBERT further introduces two new structure-aware pre-training tasks. The first one edge prediction is designed to predict whether two nodes in the data flow are connected. The other node alignment is designed to align edges between code tokens and nodes. GraphCodeBERT utilizes NL-PL pairs for six programming languages from CodeSearchNet~\cite{husain2019codesearchnet} for pre-training. It is fine-tuned on four downstream tasks including code search, clone detection, code translation and code refinement. The extensive experiments on these tasks confirm that code structures and the defined pre-training tasks help the model achieve state-of-the-art performance on these tasks.

\section{Approach}
\label{sec-app}
In this section, we first present an overview of our approach, then detail each component including PL-NL augmentation, model design in pre-training and the fine-tuning settings for downstream tasks.
\subsection{Overview}
% Figure~\ref{fig:overview} presents the overview of ContraBERT, which consists of the pre-training phase and fine-tuning phase. At the pre-training phase, it contains two components: 1) PL-NL Transformations that construct multiple semantically equivalent or similar data including the code and the comment;
% % perform the transformation operators on the program function and its comment to obtain different versions of itself. 
% 2) Pre-training, which trains the pre-trained model with the original data (i.e., functions and comments) and the new mutated versions. ContraBERT utilizes two different objectives to learn the contextual representations: MLM for predicts the original tokens in the (function, comment) pair that are masked out and InfoNCE for clustering the similar samples while pushing apart the dissimilar samples. At the fine-tuning phase, when ContraBERT is pre-trained on the unlabeled data, we fine-tune it to different types of tasks such as retrieval tasks, classification tasks, generation tasks with the task-specific labeled data with a supervised manner. 

\begin{figure}[t]
     \centering
     \includegraphics[width=0.49\textwidth]{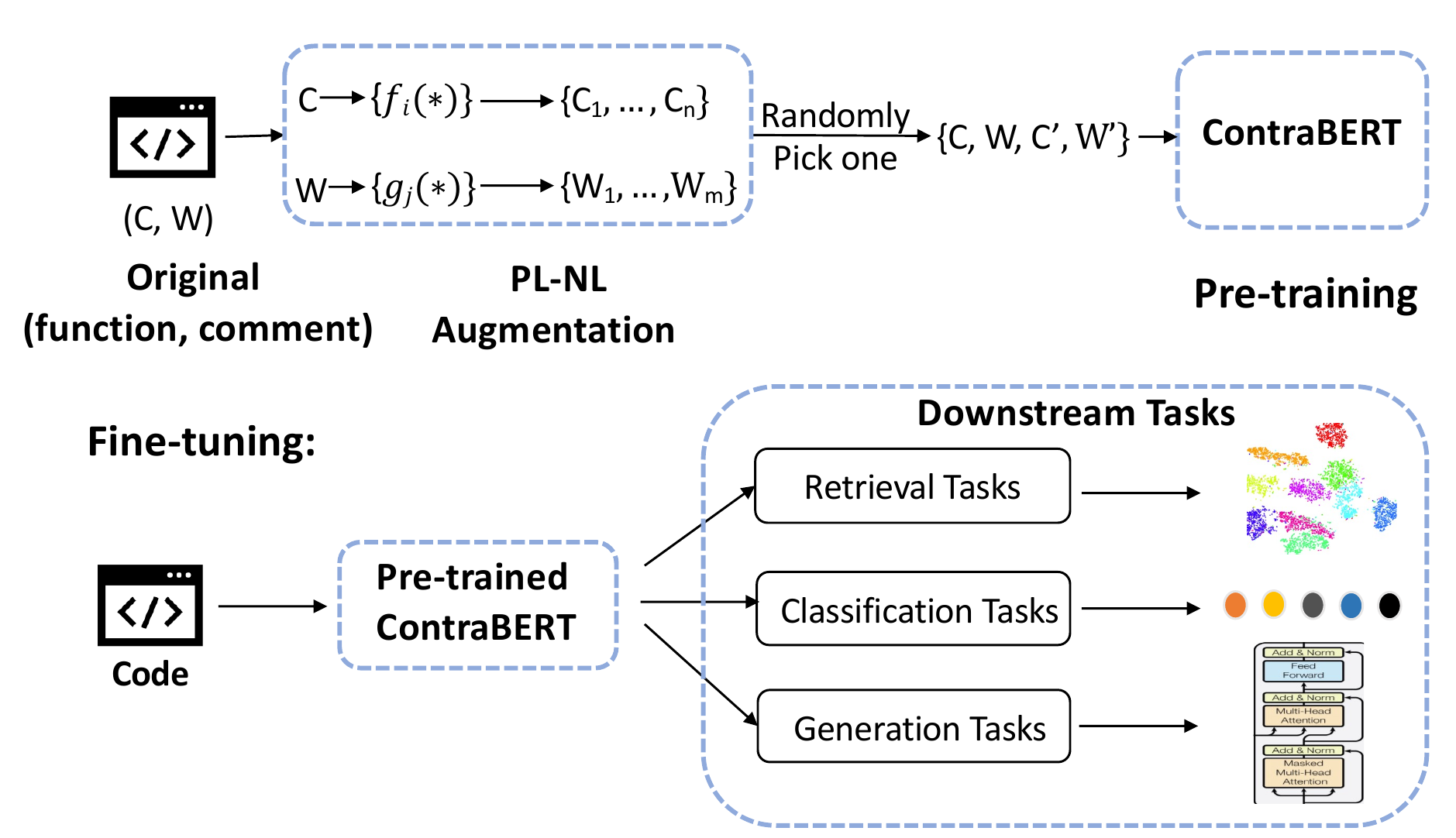}
     \caption{The Overview of ContraBERT.}
     \label{fig:overview}
\end{figure}

The overview of ContraBERT is shown in Fig.~\ref{fig:overview}. Specifically, given a pair of the function $C$  with its comment $W$ (i.e., $(C,W)$), we first design a set of PL-NL augmentation operators $\{f(*)\},\{g(*)\}$ to construct the simple or complex variants for $C$ and $W$ respectively. In the pre-training phase, initialized from existing pre-trained models such as CodeBERT or GraphCodeBERT, we further pre-train these models on the original samples and their augmented variants with masked language modeling (MLM) and contrastive pre-training task to enhance the model robustness. Finally, when ContraBERT is pre-trained over a large amount of unlabeled data, we fine-tune it for different types of tasks such as retrieval tasks, classification tasks and generation tasks with the task-specific data in a supervised manner.

\subsection{PL-NL Augmentation}\label{sec:transformation}
Given a program $C$, clone detection~\cite{svajlenko2014towards} could help to identify a semantically equivalent program $C'$. However, this technique is unrealistic in practice. For any function in a fixed dataset, we cannot guarantee that we will be able to find the semantically equivalent variants. Furthermore, clone detection usually takes a project for analysis, which is not applicable to a single function. Hence, we consider constructing augmented variants based on the original samples. Compared with the existing works~\cite{bui2021self, jain2020contrastive} that only focus on program mutations, we design a set of natural language (NL) sequence augmented operators. Specifically, we design a series of simple operators and complex operators for both PL and NL to construct variants. 
% We introduce the semantically close operators since these operations can serve as regularizers to avoid overfitting~\cite{jain2020contrastive}. 

% In the real-word, it is usually difficult or unrealistic to collect the programs that share similar semantics. In general, the functional equivalence is undecidable~\cite{rice1953classes}. Test cases~\cite{massalin1987superoptimizer} can be used to compare the programs approximately, however, this process is costly and unrealistic especially in the ``big code'' era.
% or comments are close or equivalent, especially in the ``big code'' era, %the diversity and complexity of programs and the comments in the massive dataset is unpredictable. 
\subsubsection{Program (PL) Augmentation Operators}
For program augmented operators, we design four kinds of complex operators and one kind of simple operator. 

\noindent \textbf{Complex Operators:}
\begin{itemize}[leftmargin=*]
    \item Rename Function Name (RFN). It is designed to replace the function name with a new name that is taken randomly from an extra vocabulary set constructed on the pre-training dataset. We extract all function names in the pre-training dataset for the construction. 
    Since each sample in the dataset is a single function, the renamed function preserves the equivalent semantics to the original function. 
    \item Rename Variable (RV). It renames variables in a function. A random number of variables for all occurrences in the function will be replaced with the new names taken randomly from an extra vocabulary set. We extract all variable names from the pre-training dataset to construct this vocabulary set. This operator only mutates the variable names and all occurrences of them with the new variable names, which does not change the semantics of the original function.
    \item Insert Dead Code (IDC). It means to insert unused statements in a function. To generate unused code statements, we traverse AST to identify the assignment statements and then randomly select one assignment statement to rename its variables with new names that have never appeared in the same function. After that, we consider it as the dead code and insert it at the position after the original assignment statement. As the inserted dead code does not change the original program behaviour, IDC is regarded as the semantically equivalent operator.
    \item Reorder (RO). It randomly swaps two lines of statements that have no dependency on each other in a basic block in a function body such as two declaration statements appearing on two consecutive lines without other statements between them. We traverse AST and analyze the data dependency for extraction. Since the permuted statements are independent without data dependency, this operator preserves the original program semantics.
\end{itemize}

\noindent \textbf{Simple Operators:}
\begin{itemize}[leftmargin=*]
    \item Sampling (SP). It randomly deletes one line statement from a function body and preserves others. It can serve as regularizers to avoid overfitting~\cite{jain2020contrastive}. 
\end{itemize}

\subsubsection{Comment (NL) Augmentation Operators}
Apart from the program augmentation, we further design one kind of complex operator and three kinds of simple operators for comment augmentation operators as follows:
% By parsing the original program into an abstract syntax tree (AST) for the analysis to ensure the operators maintaining the program semantics, it is harder to generate the semantic equivalent variants with the original comment in a natural language format with simple mutations. To simplify the comment mutation operator, we defined a set of semantic-close operators as follows:

\noindent \textbf{Complex Operators:}
\begin{itemize}[leftmargin=*]
    \item Back Translation Mutation (Trans). It refers to translating a source sequence into another language (target sequence) and then converting this target sequence to the original sequence~\cite{sennrich2015improving}. We use the released tool~\cite{ma2019nlpaug} for the implementation where the source is in English and the target is in German.
\end{itemize}
\noindent \textbf{Simple Operators:}
\begin{itemize}[leftmargin=*]
    \item Delete. It randomly deletes a word in a comment.
    \item Switch. It randomly switches the positions of two words in a comment.
    \item Copy. It randomly copies a word and inserts it after this word in a comment.
\end{itemize}

Given a function $C$ with its paired comment $W$, we utilize the above augmentation operators on $C$ and $W$ respectively to obtain the augmentation sets, which are defined as $S_C$ and $S_W$ respectively. Specifically, each operator is conducted once to get its corresponding augmented variant and insert it into the corresponding augmentation set. For the operator IDC, which may not get its variant for some specific functions, we ignore it and use other operators for the construction. Then we randomly select an augmented version from $S_C$ and $S_W$ (i.e., $C'\in S_C$ and $W'\in S_W$) and construct the quadruple $(C,W,C',W')$ for the pre-training. Note that during the pre-training process, at each learning step, $(C',W')$ is randomly selected from the augmented sets $S_C$ and $S_W$ respectively. Hence, each augmented sample in the sets is used when the model has sufficient learning steps. 
% which can be expressed as $S_C = \{C_i | C_i = f_i(C) \}$ and $S_W = \{W_j | W_j = g_j(W) \}$, where $f_i(*)$/$g_j(*)$ is the $i$/$j$-th operator in program/comment mutation operators. $S_C$ and $S_W$ are the obtained mutated set of variants for $C$ and $W$, respectively. 

\subsection{Model Design and Pre-training}\label{sec:contrabert}
\begin{figure*}[t]
     \centering
     \includegraphics[scale=0.45]{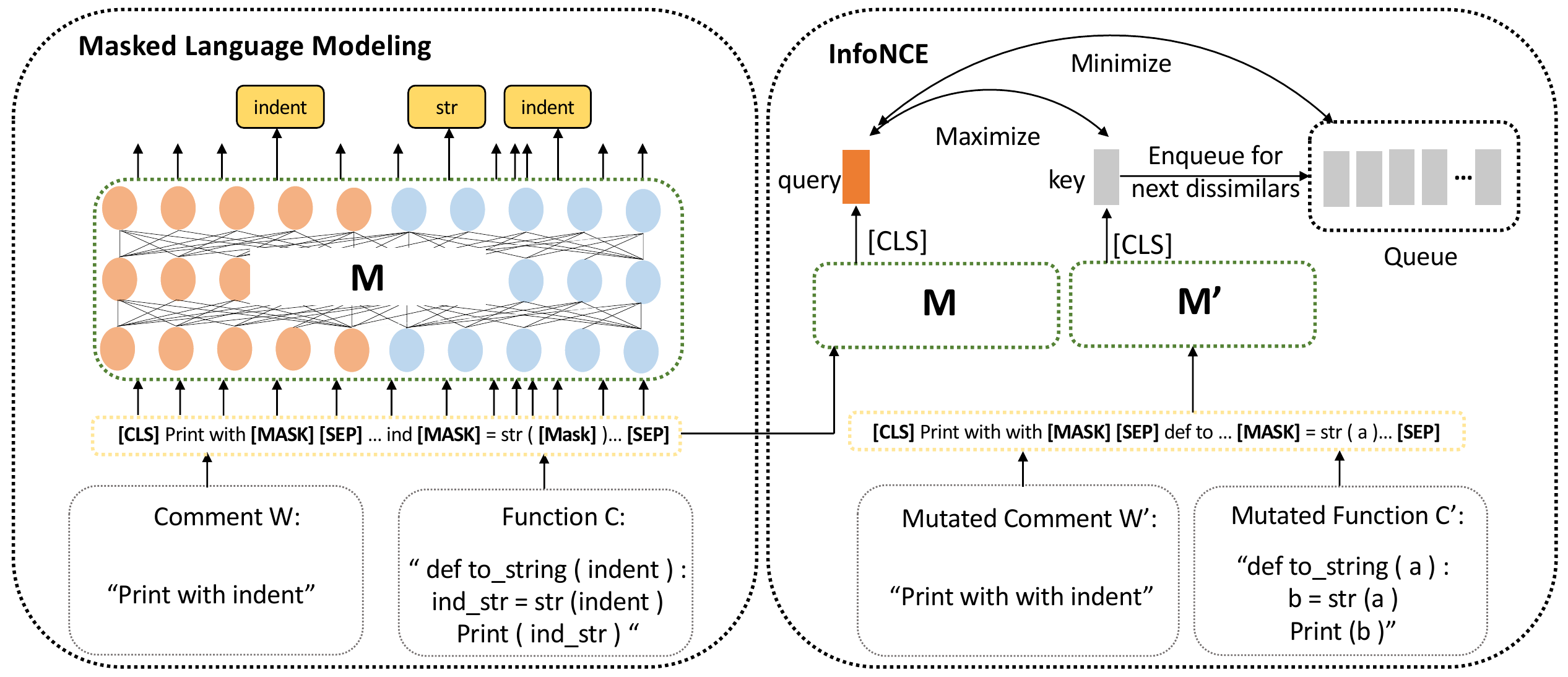}
     \caption{The model design for ContraBERT where the encoder M can be represented by the existing pre-trained models such as CodeBERT. The initial weights of the encoder M' are the same as the encoder M while the weight update is different.}
     \label{fig:arch}
\end{figure*}

Basically, ContraBERT is further trained from existing pre-trained models. We directly utilize the existing pre-trained model and further pre-train it with masked language modeling (MLM) and contrastive pre-training task to enhance its robustness. The model design of ContraBERT is presented in Fig.~\ref{fig:arch}. 
% on the original dataset with its transformed variants to shape a better representation space to improve the robustness of the original pre-trained model. The model design of ContraBERT is presented in Figure~\ref{fig:arch}. 

\subsubsection{Model Design}\label{sec:model_arc}
% Inspired by a recent finding of Jung et al.~\cite{jung2021commitbert}, utilizing CodeBERT as the initial weight has a better performance than random initialization or RoBERTa on code domain, we also leverage CodeBERT as the backbone. 

As shown in Fig.~\ref{fig:arch}, ContraBERT consists of two separate encoders M and M', where M can be represented by any pre-trained models such as CodeBERT. The model architecture of M' is the same with the encoder M and the initial weights are also the same with M. However, the weight update strategy is different with M. Specifically, given a quadruple $(C,W,C',W')$ from Section~\ref{sec:transformation}, we construct two input sequences $X=\{[CLS], W, [SEP], C, [SEP]\}$ and $X'=\{[CLS], W', [SEP], C', [SEP]\}$, where ``$[CLS]$'' indicates the beginning of a sequence and ``$[SEP]$'' is a symbol that concatenates two kinds of sequence. We utilize the encoder M and M' to encode the masked input sequence $X$ and $X'$ respectively. 

\subsubsection{Pre-training Tasks}\label{sec:pre-train-obj}
Masked language modeling (MLM) is an effective and widely adopted pre-training task to learn the effective token representations~\cite{devlin2018bert, liu2019roberta}, we also utilize it as one of our pre-training tasks. However, by our preliminary results in Section~\ref{sec-intro}, we observe that the models trained by MLM are weak to the adversarial examples, we further introduce a contrastive pre-training task to group the similar data and push away the dissimilar data to reshape the learnt space for encoder M to enhance the model robustness. 

\noindent \textbf{Masked Language Modeling (MLM).}
We utilize MLM to learn token representations in a sequence. Specifically, given the sequence $X=\{[CLS], W, [SEP], C, [SEP]\}$, a random set of positions in X are masked out. We select 15\% tokens to mask out and obtain the masked token set. Furthermore, we replace 80\% of the masked tokens in this set with the ``[MASK]'' symbol, 10\% with the random tokens from the vocabulary set and the remaining 10\% unchanged. We configure these settings since they are confirmed effective to learn the token representations in a sequence~\cite{devlin2018bert, liu2019roberta}. The loss function $\mathcal{L}_\mathrm{MLM}$ can be expressed as follows:

\begin{equation}
    \mathcal{L}_\mathrm{MLM} = -\sum_{x_i \in M} {\mathrm{log}p(x_i | X^{mask})}
\end{equation}
where $X^{mask}$ is the masked input sequence and $M$ is the masked token set. 

\noindent \textbf{Contrastive Pre-training.}
We design a contrastive pre-training task that uses InfoNCE~\cite{oord2018representation} as the loss function to enhance model robustness. It can be expressed as follows:
\begin{equation}
    \mathcal{L}_\mathrm{InfoNCE} = -\mathrm{log} \frac{\mathrm{exp}(\boldsymbol{q} \cdot \boldsymbol{k}_+/t)}{ \mathrm{exp}(\boldsymbol{q} \cdot \boldsymbol{k}_+/t) + \sum_{i=1}^n \mathrm{exp}(\boldsymbol{q} \cdot \boldsymbol{k}_i/t)}
\end{equation}
where $t$ is a temperature hyper-parameter~\cite{wu2018unsupervised}, the query vector $\boldsymbol{q}$ is the encoded vector representation, $\boldsymbol{k}_+$ is a similar key vector that $\boldsymbol{q}$ matches, $\boldsymbol{K} = \{\boldsymbol{k}_1, ..., \boldsymbol{k}_n\}$ is a set of dissimilar encoded vectors. %InfoNCE tries to classify the query $\boldsymbol{q}$ as the similar sample $\boldsymbol{k}_+$ and pushes away the dissimilar samples in $K$. 
InfoNCE tries to classify the query vector $\boldsymbol{q}$ into its similar sample $\boldsymbol{k}_+$ and pushes it away from dissimilar samples in the set $\boldsymbol K$. The similarity is measured by dot product $(\cdot)$ between two vectors. 
To obtain the query representation $\boldsymbol{q}$ and the similar key representation $\boldsymbol{k}_+$, inspired by the recent advance~\cite{he2020momentum} on the image recognition, we adopt Momentum Contrast (MoCo)~\cite{he2020momentum} for the encoding. Specifically, it introduces an extra encoder M' to get the key representation $\boldsymbol{k}_+$, which can be expressed as follows:
\begin{equation}
\label{eq:contra}
\begin{split}
    \begin{gathered}
    \boldsymbol q =\mathrm{LayerNorm}(\mathrm{M}(X)[0]) \\ 
    \boldsymbol k_+ =\mathrm{LayerNorm}(\mathrm{M}'(X')[0])
    \end{gathered}
\end{split}
 \end{equation}
where $X$ and $X'$ denote the original masked sequence and its mutated variant respectively. The index 0 denotes the position of ``[CLS]'' in the sequence, which can be considered as the aggregated sequence representation. The encoder M' is the same as the encoder M, but during the learning phase, it utilizes a momentum to update its learnt weights while the encoder M uses the gradient descent:

\begin{equation}
\theta_{\mathrm{M'}}\leftarrow m\theta_{\mathrm{M'}} + (1-m)\theta_{\mathrm{M}}
\end{equation}
where $m \in [0, 1)$ is a momentum coefficient for scaling, $\theta_{\mathrm{M'}}$ and $\theta_{\mathrm{M}}$ denote the learnt weights for model $\mathrm{M'}$ and M.

From Eq~\ref{eq:contra}, we obtain the query representation $\boldsymbol{q}$ and the key representation $\boldsymbol{k}_+$, to compute the similarity with dissimilar vectors from $\boldsymbol{K}$, MoCo maintains a ``\textit{dynamic}'' \textit{queue} of length $n$. This queue stores the dissimilar keys from the previous batches. Specifically, during the learning phase, the current query $\boldsymbol q$ will calculate the similarity with all dissimilar vectors in this queue. Afterwards, the key vector $\boldsymbol{k}_+$ will be enqueued to \textit{queue} to replace the oldest one and we take it as the dissimilar samples for the calculation of the next query. Hence, it is namely \textit{dynamically updated}.

Finally, we add both loss values with the scaled factor to pre-train ContraBERT and this process is expressed as follows:

\begin{equation}
\label{eq:loss}
\mathcal{L}_\mathrm{Loss} = \mathcal{L}_\mathrm{MLM} + w  \mathcal{L}_\mathrm{InfoNCE}
\end{equation}
where $w$ is the hyper-parameter to scale the weight for both pre-training tasks.  

\subsection{Fine-tuning}
Once ContraBERT is further pre-trained from the original pre-trained model, we can utilize it to obtain the vector representation for a program. Furthermore, we can also transfer it to different downstream tasks during the fine-tuning phase. 
%we roughly categorized into three different types of tasks i.e., retrieval tasks e.g., clone detection~\cite{mou2016convolutional, wang2020detecting}, code search~\cite{gu2018deep, husain2019codesearchnet}; classification tasks e.g., defect-detection~\cite{zhou2019devign}; generation tasks e.g., code-to-code translation~\cite{chen2018tree, lachaux2020unsupervised}, code-refinement~\cite{tufano2019empirical}. 
These downstream tasks can be roughly categorized into three groups: (1) retrieval tasks (e.g., clone detection~\cite{mou2016convolutional, wang2020detecting}, code search~\cite{gu2018deep, husain2019codesearchnet}); (2) classification tasks (e.g., defect-detection~\cite{zhou2019devign}); (3) generation tasks (e.g., code-to-code translation~\cite{chen2018tree, lachaux2020unsupervised}, code-refinement~\cite{tufano2019empirical} and source code summarization~\cite{liu2020retrieval}). Since the output space may differ from the pre-trained space, similar to CodeBERT and GraphCodeBERT, we add the task-specific module and then fine-tune the completed network on the labeled data. Specifically, for retrieval tasks, we further train ContraBERT on a labeled dataset; for classification tasks, we add a multi-layer perceptron (MLP) to predict the probability for each class; for generation tasks, we add a Transformer-based decoder to generate the target sequence.

\section{Experimental Setup}
\label{sec-experimentalsetup}
%In this section, we first introduce a list of downstream tasks with the corresponding datasets to evaluate, then present each metric where the task uses as well as the experimental settings. 

In experiments, we first evaluate the effectiveness of our approach (RQ1) in improving model robustness. Then we plot the feature space learnt by different pre-trained models for visualization to confirm the features are learnt better(RQ2). Finally, we conduct extensive experiments to demonstrate the robustness-enhanced models provide significant improvements on downstream tasks (RQ3-RQ4). The detailed research questions are described as follows:

\begin{itemize}[leftmargin=*]
\item {\bf RQ1:} {What is the performance of different augmentation operators in enhancing the robustness of the pre-trained model?} 
\item {\bf RQ2:} {Can ContraBERT reshape the vector space learnt from the pre-trained models to obtain better vector representations? } 
\item {\bf RQ3:} {Can ContraBERT outperform the original pre-trained models on different downstream tasks?} 
\item {\bf RQ4:} {Are the defined pre-training tasks both effective in improving the downstream task performance?}
% \item {\bf RQ5:} {What is the impact of different operators in PL-NL mutations on the performance of the downstream tasks?}
\end{itemize}

% In this section, we first briefly introduce a list of downstream tasks with the corresponding used datasets for evaluation, then present each evaluation metric for different tasks followed by the experimental settings for ContraBERT.

\subsection{Evaluation Tasks, Datasets and Baselines}
We select four downstream tasks for evaluation. They are clone detection~\cite{mou2016convolutional, svajlenko2014towards}, code search~\cite{husain2019codesearchnet}, defect detection~\cite{zhou2019devign} and code translation~\cite{chen2018tree, lachaux2020unsupervised}. We briefly introduce each task as follows:

\noindent \textbf{Clone Detection (Code-Code Retrieval).}
This task is to identify semantically equivalent programs from a set of distractors by measuring the semantic similarity between two programs. AI for clone detection calculates cosine similarity between two embedding vectors of programs produced by neural networks and selects the top-k most similar programs as the candidates.

\noindent \textbf{Defect Detection (Code Classification).}
It aims to detect whether a function contains defects that will be exploited to attack the software systems. %Due to that the defect in the program is hard to detect by the traditional techniques, recent advanced works~\cite{zhou2019devign, li2018vuldeepecker, russell2018automated} propose to employ deep neural network to learn program semantics to facilitate the detection.
Since the defects in a program are still difficult to be effectively detected by the traditional techniques, recently advanced works~\cite{zhou2019devign, li2018vuldeepecker, russell2018automated} propose to employ a deep neural network to learn program semantics to facilitate the detection. These AI-based techniques predict the probability of whether a function is vulnerable or not.
% We utilize the defect detection dataset, provided by CodeXGLUE for evaluation. 

\noindent \textbf{Code Translation (Code-Code Generation).}
%This aims to translate a language of program e.g., Java to a semantic equivalent one in another language e.g., C\#.
It aims to translate a program in a programming language (e.g., Java) to the semantically equivalent one in another language (e.g., C\#).
Some previous works~\cite{lachaux2020unsupervised} analogy it to machine translation~\cite{vaswani2017attention, bahdanau2014neural} in NLP and employ LSTMs~\cite{hochreiter1997long} and Transformer~\cite{vaswani2017attention} for code translation. 

% \noindent \textbf{Code Refinement (Code-Code Generation).}
% The goal of code refinement is to fix the bug in the program automatically. Specifically, given a buggy function, it targets outputting a refined function with the bug fixed. Similar to code translation, recent advances~\cite{tufano2019empirical} formulate this task as the generation task in NLP and utilize LSTMs~\cite{hochreiter1997long} and the Transformer~\cite{vaswani2017attention} for the automated patching.

\noindent \textbf{Code Search (Text-Code Retrieval).}
It aims at returning the desired programs based on the query in a natural language. Similar to clone detection, it measures the semantic relevance between queries and programs. The input for the deep code search system~\cite{husain2019codesearchnet, gu2018deep} is a natural language query and the output is programs that meet the query requirements. The cosine similarity is used to compute semantic similarity between the vectors of a query and programs.

% \subsubsection{Code Summarization (Code-Text)}
% The objective of source code summarization is to generate the natural language comment for a function to describe the functionality of the program. Similar to code translation,  LSTM~\cite{hochreiter1997long} and Transformer~\cite{vaswani2017attention} are widely adopted in this task. 

% Besides the above evaluation tasks, there are some other tasks in CodeXGLUE i.e., code completion~\cite{allamanis2013mining, raychev2016probabilistic}, text-to-code generation~\cite{iyer2018mapping}. For these tasks, CodeXGULE utilized GPT-2 as the architecture, which consists of 12 layers Transformer decoder and the model architecture is different from CodeBERT, we neglect these tasks and leave them as our future work. 

In terms of the pre-training dataset, we use the released dataset provided by CodeSearchNet~\cite{husain2019codesearchnet} and this dataset is also used by CodeBERT and GraphCodeBERT. We use bimodal NL-PL pairs for pre-training, which consist of six programming languages including Java, Python, Ruby, Go, PHP and JavaScript. For the fine-tuning datasets, for the tasks of clone detection (POJ-104), defect detection, and code translation, we directly utilize the released task-specific dataset provided by CodeXGLUE~\cite{lu2021codexglue}. For code search, we use the cleaned dataset provided by GraphCodeBERT~\cite{guo2020graphcodebert} for evaluation. For each task, we utilize the official scripts to make a fair comparison. In addition, by the defined augmentation operators in Section~\ref{sec:transformation}, we obtain a large amount of extra data $(C', W')$ used in \tool as compared to the original pre-training data used in CodeBERT and GraphCodeBERT. Hence, we further add two baselines CodeBERT\_Intr and GraphCodeBERT\_Intr, which utilize original data as well as the dataset of the extra data $(C', W')$ to pre-train CodeBERT and GraphCodeBERT with MLM for comparison.

\subsection{Evaluation Metrics}
In ContraBERT, different metrics are used to evaluate downstream tasks. We follow the metrics that CodeXGLUE used for evaluation, and the details are listed below:

% , defined as follows:
% \begin{equation}
%   \mathrm{MRR} = \frac{1}{|Q|} \sum_{q=1}^{|Q|} \frac{1}{\mathrm{FRank}_q}
% \end{equation}
% where $Q$ is a set of queries, $\mathrm{FRank}_q$ is the rank position of the first hit for the query $q$ in the result list. 
%Instead of retrieving 1,000 candidates that CodeBERT used~\cite{feng2020codebert}, we follow GraphCodeBERT settings~\cite{guo2020graphcodebert}, to retrieve the answer of each query from the whole test set i.e., the result list size is equal to the size of test set. %MRR is used to evaluate the performance of code search. 

\noindent \textbf{MAP@R.}
%MAP@R is the abbreviation of the mean of average precision and it is evaluated for retrieving R most similar samples given a query. MAP@R is utilized in clone detection, where R is set to 499 for evaluation. 
It is the abbreviation of the mean of average precision, which is used to evaluate the result of retrieving R most similar samples in a set given a query. MAP@R is used for clone detection, where R is set to 499 for evaluation. 

\noindent \textbf{Acc.}
It defines the ratio of correct predictions (i.e., the exact match) in the testset. Acc is used for the evaluation of defect detection and code translation. 

\noindent \textbf{BLEU-4.}
%BLEU-4 is widely used to evaluate the text similarity of the generated sequence with the ground-truth. In ContraBERT, we utilize BLEU-4 for the evaluation on code translation and code refinement. 
It is widely used to evaluate the text similarity between the generated sequence with the ground-truth in the generation systems. We use BLEU-4 for code translation.

% \noindent \textbf{CodeBLEU.}
% CodeBLEU~\cite{ren2020codebleu} is a newly designed metric to assess the quality of generated code snippets. Different from BLEU-4 that only computes the text similarity, CodeBLEU further incorporates code syntax and code semantic information to improve the evaluation accuracy. CodeBLEU is used to evaluate code translation and code refinement. 

\noindent \textbf{MRR.}
%MRR is the abbreviation of mean reciprocal rank and it is widely adopted in information retrieval systems~\cite{gu2018deep, wang2009portfolio}. MRR can be obtained by the inverse of the rank of the first hit result in the result list.
It is the abbreviation of Mean Reciprocal Rank, which is widely adopted in information retrieval systems~\cite{gu2018deep, wang2009portfolio}. 
% It can be obtained by the inverse of the ranking of the first hit result in the result list. 
We used it to evaluate the performance of code search. Instead of retrieving 1,000 candidates like CodeBERT~\cite{feng2020codebert}, we follow the settings of GraphCodeBERT~\cite{guo2020graphcodebert} to retrieve the answer for each query from the whole test set.

%Overall, we can conclude that the higher values of these metrics, the better performance of the approach it achieves. 
% Overall, we can conclude that the higher values of these metrics, the better performance the approach achieves.

\subsection{Experimental Settings}\label{sec:settings}
We adopt CodeBERT and GraphCodeBERT as our original models. We set the maximum input sequence length $X$ and the mutated sequence $X'$ as 512 following CodeBERT. We use Adam for optimizing with 256 batch size and 2e-4 learning rate. %At each iteration, $X'$ is constructed with $C'$ and $W'$ randomly picked from $S_C$ and $S_W$.
At each iteration, $X'$ is constructed by $C'$ and $W'$, which are randomly picked from $S_C$ and $S_W$ respectively.
Following He et al.~\cite{he2020momentum}, the momentum coefficient $m$, temperature parameter $t$ and \textit{queue} size is set to 0.999, 0.07 and 65536 accordingly. We set the weight $w$ in Eq~\ref{eq:loss} as 0.5 to accelerate the coverage process. The model is trained on a DGX machine with 4 NVIDIA Tesla V100 with 32GB memory.
%To alleviate the bias towards the high-resource languages i.e., the number of samples for different programming language is different, following Guo et al.~\cite{guo2020graphcodebert}, we sample each batch from the same programming language according to a multinomial distribution with probabilities ${\{q_i\}}_{i=1...N}$~\cite{lample2019cross}.
To alleviate the bias towards the high-resource languages (i.e., the number of samples for different programming languages is different), we refer to GraphCodeBERT~\cite{guo2020graphcodebert} and sample each batch from the same programming language according to a multinomial distribution with probabilities ${\{q_i\}}_{i=1...N}$.
\begin{equation}\label{sample}
q_i=\frac{p^\alpha_i}{\sum_{j=1}^{N}p^\alpha_j} \ with \ \ p_i=\frac{n_i}{\sum_{k=1}^{N}n_k}
\end{equation}
where $n_i$ is the number of samples for $i$-th programming language, $N$ is the total number of languages and $\alpha$ is set to 0.7. The model is trained with 50K steps to ensure each mutated sample is utilized for the learning process and it takes about 2 days to finish the pre-training process. %At the fine-tuning phase, for the settings of all downstream tasks, we directly utilize the default settings from CodeXGLUE~\cite{lu2021codexglue} for fairness. 
At fine-tuning phase, we directly utilize the default settings of CodeXGLUE~\cite{lu2021codexglue} and GraphCodeBERT~\cite{guo2020graphcodebert} in ContraBERT for downstream tasks. All experiments of downstream tasks are conducted on Intel Xeon Silver 4214 Processor with 6 NVIDIA Quadro RTX 8000 with 48GB memory.

\section{Experimental Results}
\label{sec-evaluation}

\begin{table*}[t]
\centering
\scriptsize
\caption{Results of ContraBERT against the variable renaming operator in a zero-shot manner.}
\label{tbl-ablation-adversarial}
\begin{tabular}{lccccc|lccccc}\hline
 \multirow{2}{*}{Model}  & \multirow{2}{*}{Num} & N=0 &N=1    &N=4    &N=8 & \multirow{2}{*}{Model}  & \multirow{2}{*}{Num} & N=0 &N=1  &N=4  &N=8   \\
     &  & Acc & Acc  & Acc & Acc  &&   & Acc & Acc  & Acc & Acc   \\ \hline
ContraBERT\_C w/o RFN          & 10,087  & 1 & 0.977  & 0.868  & 0.634    & ContraBERT\_G w/o RFN        & 10,375 & 1 & 0.975  &0.873 &  0.634\\
ContraBERT\_C w/o RV           & 8,665   & 1 & 0.932  & 0.597  & 0.291  & ContraBERT\_G w/o RV           &  9,042 & 1 & 0.955  & 0.657 & 0.309 \\
ContraBERT\_C w/o IDC          & 9,997   & 1 & 0.969  & 0.865  & 0.618  & ContraBERT\_G w/o IDC          & 10,530  & 1 & 0.963 & 0.862 & 0.612 \\
ContraBERT\_C w/o RO           & 9,923  & 1 & 0.963  & 0.857  & 0.619   & ContraBERT\_G w/o RO           & 10,509 & 1 & 0.968 & 0.868  & 0.617 \\
ContraBERT\_C w/o SP           & 10,604  & 1 & 0.959  & 0.849  & 0.616  & ContraBERT\_G w/o SP           & 11,140 & 1 & 0.969 &0.860 & 0.613 \\\hline
% ContraBERT\_C w/o Syn       &  & 1 &   &   &    & ContraBERT\_G w/o Syn & & 1 &  &   & \\  
ContraBERT\_C w/o Trans       &9,536  & 1 & 0.971  & 0.856  &0.621    & ContraBERT\_G w/o Trans & 10,360 & 1 & 0.973 & 0.859  & 0.617 \\  
ContraBERT\_C w/o Delete       & 10,199 & 1 & 0.978  & 0.871  & 0.639   & ContraBERT\_G w/o Delete       & 10,376& 1 & 0.981 & 0.878  &0.643 \\   
ContraBERT\_C w/o Switch       & 9,809  & 1 & 0.975  & 0.877  & 0.637  & ContraBERT\_G w/o Switch       & 10,457 & 1 & 0.978& 0.876 &0.647 \\
ContraBERT\_C w/o Copy         & {10,749}  & 1 & 0.977  & 0.874  & 0.635 & ContraBERT\_G w/o Copy & 10,859 & 1 & 0.981 & 0.880 &0.641 \\\hline
{ContraBERT\_C}  & 10,463  & 1 & \textbf{0.981}  & \textbf{0.882} & \textbf{0.649}  & {ContraBERT\_G} & 10,565  & 1 & \textbf{0.985} & \textbf{0.888} & \textbf{0.654}\\ \hline
\end{tabular}
\end{table*}

\subsection{RQ1: Robustness Enhancement.}\label{sec:rq1}
We investigate the augmentation operators in enhancing model robustness by validating the accuracy of samples against adversarial attacks on clone detection (POJ-104). The main reason to choose clone detection is that it targets identifying the semantically equivalent samples from other distractors. Hence, although the variable renaming operator changes the text of a program, the original program semantics are still unchanged. We statistically analyse the correctly predicted results under a different number of renaming edits for illustration. The experiments are conducted in a zero-shot manner~\cite{palatucci2009zero}, which means that it does not involve fine-tuning phase and we directly utilize the pre-trained model for evaluation.
Specifically, we remove one operator and keep the remaining operators in Section~\ref{sec:transformation} to pre-train the model. 
For fairness, the other settings in the experiments are the same as ContraBERT.
Then we utilize the testset (in total 12,000 samples) on clone detection (POJ-104) and randomly mutate the variables contained in the correctly predicted samples produced by different pre-trained models from 1 to 8 edits to test the prediction accuracy. The experimental results are shown in Table~\ref{tbl-ablation-adversarial} where N is the number of edits and Num is the total number of correctly predicted samples without any edits in the testset for different models. ContraBERT\_C/G defines the model is initialized by CodeBERT and GraphCodeBERT respectively and {w/o $*$} defines the removed operator $*$.

From Table~\ref{tbl-ablation-adversarial}, we find that in general, with the increasing number of edits, the performance continues to drop. It is reasonable, as the increasing number of edits, the difficulty for corrected predictions is also increased. We also observe that each augmented operator is beneficial to improve model robustness against the adversarial samples and when incorporating all operators, we obtain the best performance. It demonstrates the effectiveness of our designed PL-NL augmentation operators. 
In terms of NL augmentation operators, the operators Delete/Switch/Copy are relatively weaker in the robustness enhancement compared with the operator Trans. Since the operators (Delete/Switch/Copy) just have a limited extent of modification on the original sequence (i.e., only one or two words are modified), the text similarity between $W$ and $W'$ is more similar than the operator Trans produces. Hence, the data diversity is limited by Delete/Switch/Copy, which leads to the robustness improvement is not as obvious as the operator Trans. 
In terms of PL augmentation operators, we find that the number of correctly predicted samples of ContraBERT\_C/G w/o RV is the lowest (e.g., 8,665 and 9,042). With the increasing number of edits, the accuracy drops by a great margin. This indicates that RV operator plays a critical role against adversarial attacks and removing it harms the performance significantly. In addition, removing RFN operator, ContraBERT also has higher accuracy than other PL operators (i.e., RV, IDC, RO and SP), which indicates that RFN has fewer contributions. It is caused by the generated program $C'$ by RFN (i.e., rename function name) is more similar to the original program $C$ compared with other PL augmentation operators. 

\begin{tcolorbox}[breakable,width=\linewidth,
boxrule=0pt,top=1pt, bottom=1pt, left=1pt,right=1pt, colback=gray!20,colframe=gray!20]
\textbf{\ding{45} $\blacktriangleright$RQ1$\blacktriangleleft$}
Each operator in the designed PL-NL augmentation is effective in improving model robustness and when incorporating them, the robustness of pre-trained models is further enhanced.
\end{tcolorbox}

\begin{figure*}
     \centering
    %  \begin{subfigure}[b]{0.3\textwidth}
    %      \centering
    %      \includegraphics[scale=0.275]{./figures/roberta_visual_test.pdf}
    %      \caption{RoBERTa}
    %      \label{fig:visual-roberta}
    %  \end{subfigure}
    %  \hfill
     \begin{subfigure}[b]{0.2\textwidth}
         \centering
         \includegraphics[scale=0.2]{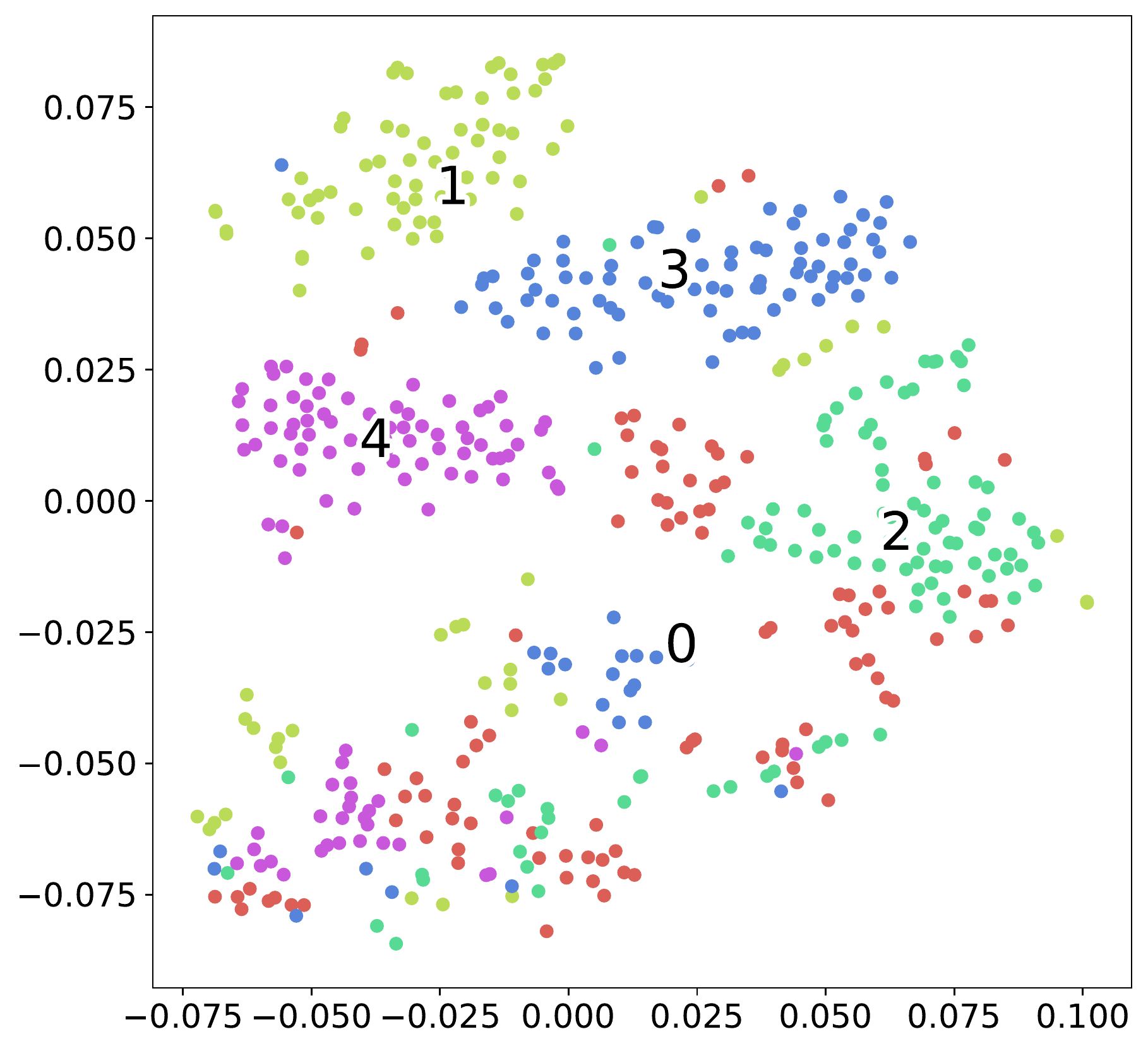}
         \caption{CodeBERT}
         \label{fig:visual-codebert}
     \end{subfigure}
     \hfill
     \begin{subfigure}[b]{0.2\textwidth}
         \centering
         \includegraphics[scale=0.2]{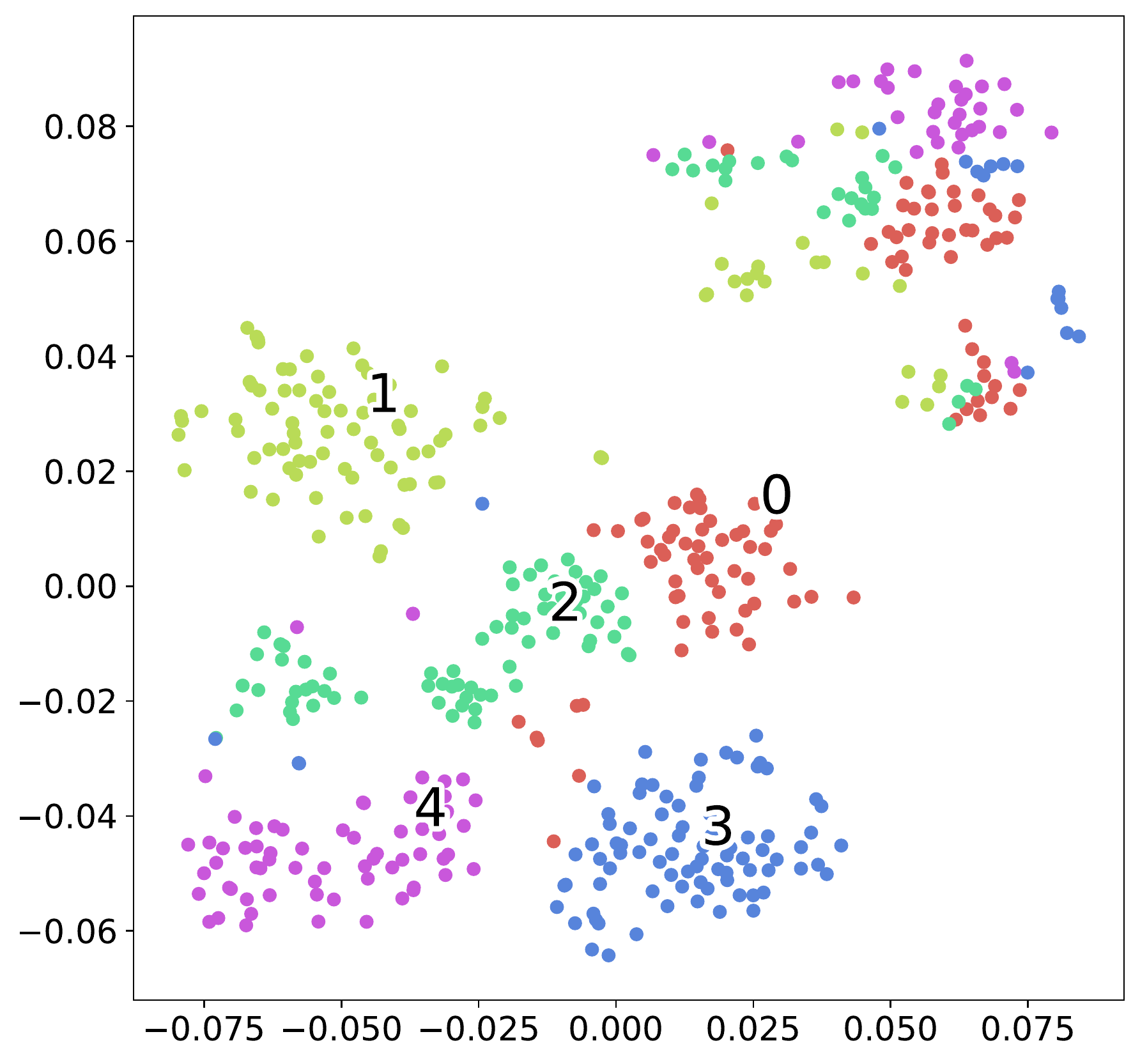}
         \caption{GraphCodeBERT}
         \label{fig:visual-graphcodebert}
     \end{subfigure}
     \hfill
     \begin{subfigure}[b]{0.2\textwidth}
         \centering
         \includegraphics[scale=0.2]{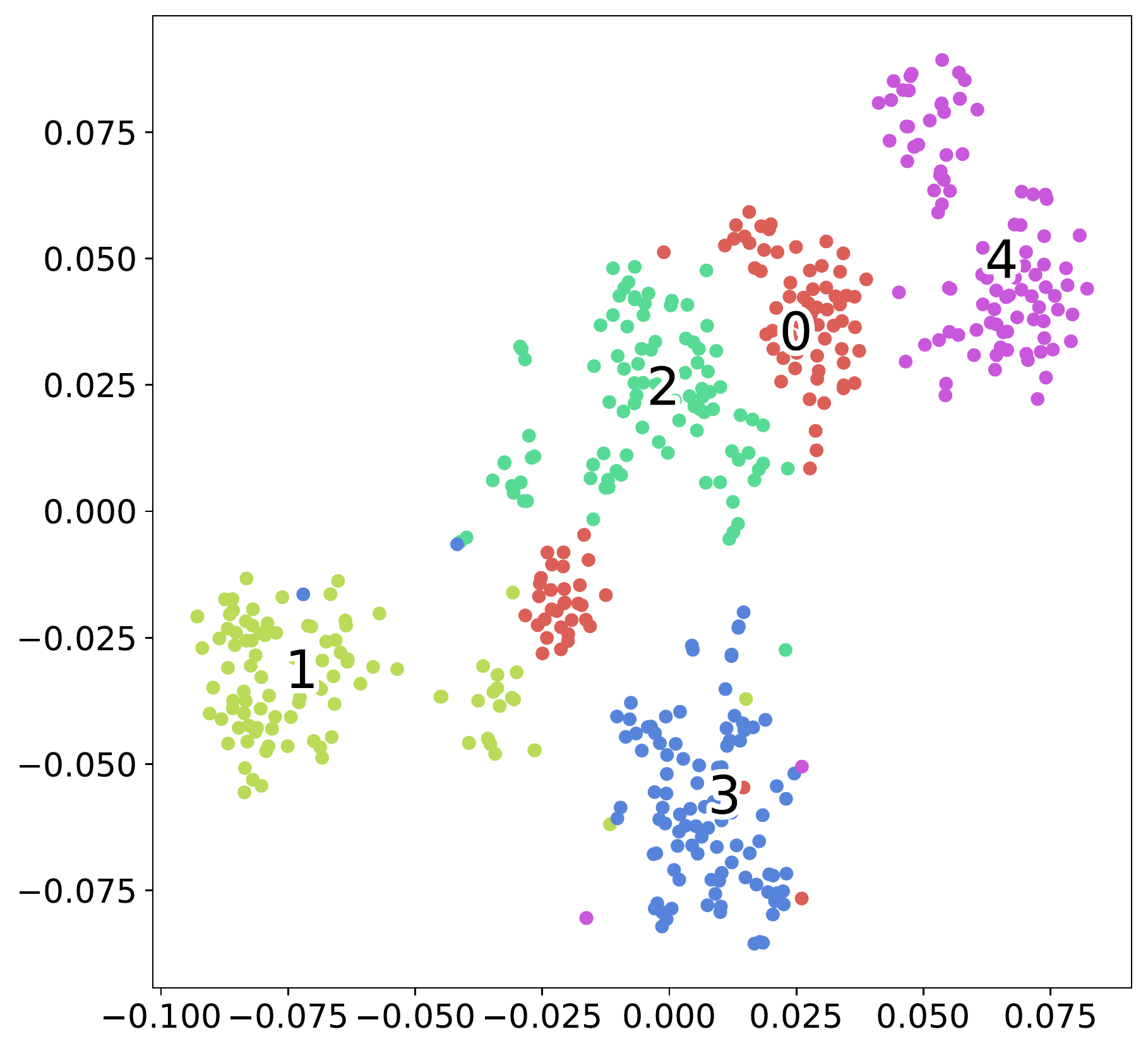}
         \caption{ContraBERT\_C}
         \label{fig:visual-contrabert-c}
    \end{subfigure}
    \hfill
       \begin{subfigure}[b]{0.2\textwidth}
         \centering
         \includegraphics[scale=0.2]{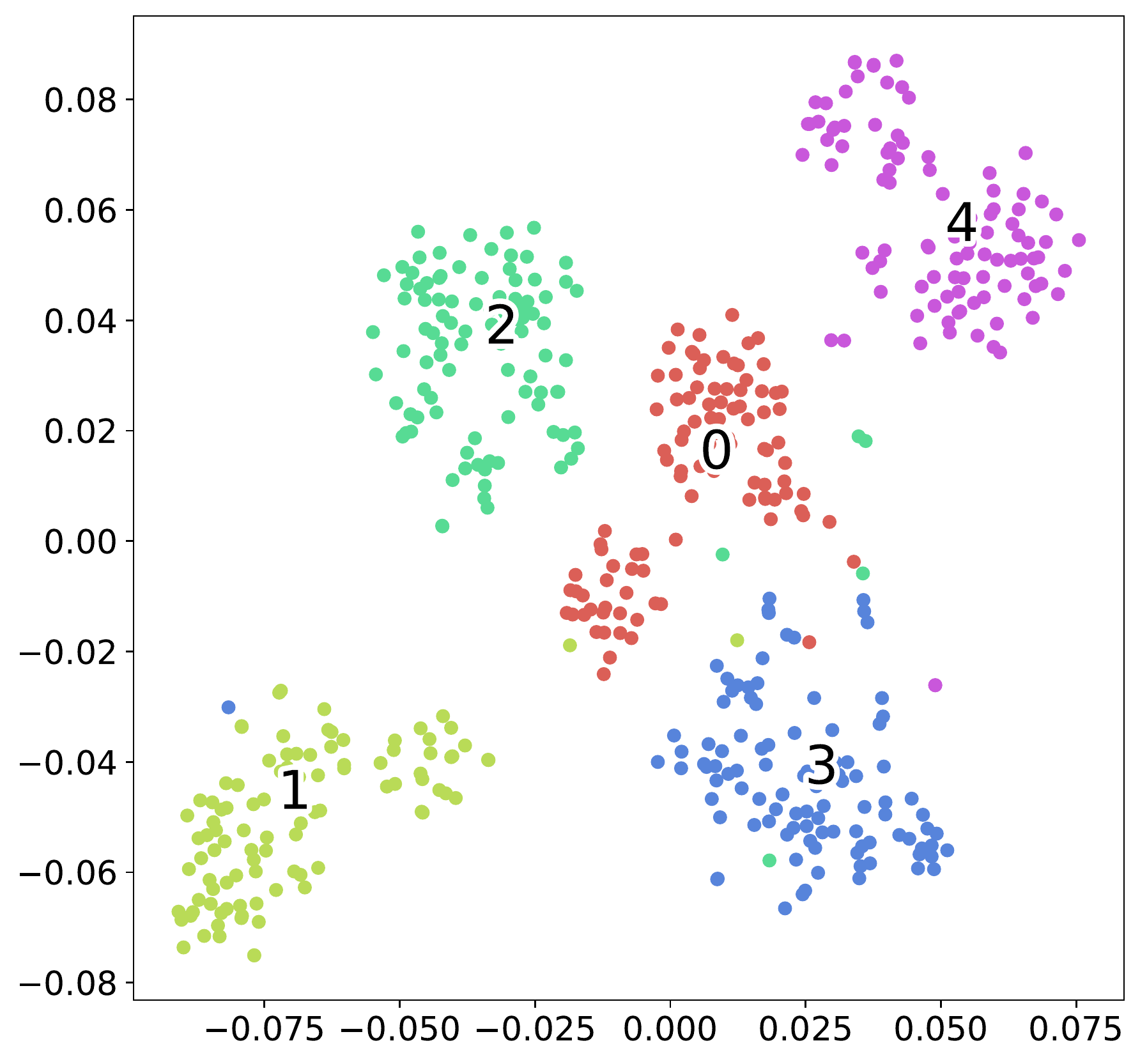}
         \caption{ContraBERT\_G}
         \label{fig:visual-contrabert-g}
     \end{subfigure}
        \caption{Visualization for vector representations of each 100 programs for 5 problems and they are randomly picked from clone detection (POJ-104). The vectors are produced by CodeBERT, GraphCodeBERT, ContraBERT\_C and ContraBERT\_G. The point with different colours indicates different problems that this function belongs to.}
         \label{fig:visual}
\end{figure*}

\subsection{RQ2: Visualization for Code Embeddings.}\label{sec:rq-visualization}
We visualize the code representation space learnt by different pre-trained models to confirm that the contrastive pre-training task can reshape the learnt vector space to ensure the model is more robust.
Specifically, we use the clone detection (POJ-104) task provided by CodeXGLUE~\cite{lu2021codexglue} for evaluation. The main reason for selecting clone detection is that it is more intuitive to observe and validate the similarity of code representation on the semantic equivalence programs. The dataset consists of 104 programming problems, where each problem has 500 semantically equivalent programs with different implementations. Theoretically, the program semantics for one problem should be the same. Hence, the code vectors (i.e., representations) of programs from pre-train models for one problem should be closer than the code vectors of programs for other problems.
% are expected to be aggregated in one cluster and different classes may have their own clusters after the neural network embeddings. Hence, a
% \wbz{Hence, theoretically, through the neural network embeddings, the programs in the same class will be aggregated into one cluster and different classes may have their own clusters.} 
We randomly select 5 different problems with 100 samples and take them as the inputs to CodeBERT, GraphCodeBERT, ContraBERT\_C and ContraBERT\_G for visualization where C/G defines ContraBERT is initialized by CodeBERT or GraphCodeBERT respectively. We utilize the vector of the token ``[CLS]'' as the program representation.  We further utilize T-SNE~\cite{van2008visualizing} to reduce the vector dimension to a two-dimensional space for visualization. 
Similar to Section~\ref{sec:rq1}, this process is also zero-shot~\cite{palatucci2009zero}, which helps us to validate the learnt space by different pre-training techniques.

As shown in Fig.~\ref{fig:visual}, the vectors produced by GraphCodeBERT (See Fig.~\ref{fig:visual-graphcodebert}) have a certain ability to group some problems of programs compared with CodeBERT (See Fig.~\ref{fig:visual-codebert}), which indicates that incorporating program structures such as data flow graph into pre-training is beneficial for the model to learn program semantics. However, we also find that the improvement is limited and
the boundary in Fig.~\ref{fig:visual-graphcodebert} is not clear. Some data points are scattered, especially in the upper-right part of Fig.~\ref{fig:visual-graphcodebert}. In contrast, the visualization of ContraBERT is shown in Fig.~\ref{fig:visual-contrabert-c} and Fig.~\ref{fig:visual-contrabert-g}. We see that the programs in the same problem aggregate together closely as a cluster and different clusters have much clearer boundaries. This indicates that ContraBERT is more powerful than CodeBERT/GraphCodeBERT to group semantically equivalent data and push away dissimilar data. We attribute this ability to the defined PL-NL augmentation operators to capture the essence of programs.
Furthermore, ContraBERT\_G (See Fig.~\ref{fig:visual-contrabert-g}) has a better clustering performance than ContraBERT\_C (See Fig.~\ref{fig:visual-contrabert-c}). For example in Fig.~\ref{fig:visual-contrabert-c}, the label 0 has two distant clusters while in Fig.~\ref{fig:visual-contrabert-g}, it only has one cluster. The improvements are from the used original model that GraphCodeBERT is superior to CodeBERT. In addition, we compute the distortion distance~\footnote{The distortion distance refers to the sum of the squared distances of each sample to their assigned cluster centre.}~\cite{Distortion} of the selected samples for these models to strengthen the conclusion. The distances of CodeBERT, GraphCodeBERT, ContraBERT\_C and ContraBERT\_G are 0.333, 0.212, 0.202, and 0.194 respectively. We can find that ContraBERT has a lower distortion distance than CodeBERT and GraphCodeBERT, which demonstrates their clusters are more compact. 

% As shown in Figure~\ref{fig:visual}, the vectors produced by CodeBERT (See Figure~\ref{fig:visual-codebert}) can , which is pre-trained on the large natural language corpus. This proves that the knowledge learnt from the natural language by RoBERTa may not be useful for programs and employing the large program corpus for the pre-training can learn a better program representations. However, we also find that the boundary in Figure~\ref{fig:visual-codebert} is not clear, especially for the class 0 (the points in red color), which is scattered. The visualization of ContraBERT is shown in Figure~\ref{fig:visual-contrabert}, we can see that the programs with the same label aggregate together as a cluster and different clusters have much clearer boundaries. 

\begin{tcolorbox}[breakable,width=\linewidth,
boxrule=0pt,top=1pt, bottom=1pt, left=1pt,right=1pt, colback=gray!20,colframe=gray!20]
\textbf{\ding{45} $\blacktriangleright$RQ2$\blacktriangleleft$} 
Through contrastive pre-training tasks to learn augmented variants constructed by a set of PL-NL operators, ContraBERT is able to group 
the semantically equivalent samples and push away the dissimilar samples, thus learning better vector representations. 
\end{tcolorbox}

\begin{table}[]
\centering
\caption{Results on clone detection and defect detection.}
\label{tbl-clone-defect}
\begin{tabular}{lcc}\hline
 \multirow{2}{*}{Model}                      &Clone Detection  & Defect Detection \\
           & MAP@R  & Acc\\ \hline
% Code2Vec            & 1.98   &-\\
% NCC                 & 54.19  &-\\
% Aroma               & 55.12  & -\\
% MISIM-GNN           & 82.45  & -\\
% BiLSTM              &  -     & 59.37    \\
% TextCNN             &  -     & 60.69    \\
% RoBERTa             & 79.96  & 61.05\\ \hline
CodeBERT            & 84.29  & 62.08 \\ 
CodeBERT\_Intr  & 86.34&62.41\\
{ContraBERT\_C} (MLM) &86.21 & 62.25 \\
{ContraBERT\_C} (Contra) &  81.44 &  62.22 \\
{ContraBERT\_C}           & \textbf{90.46}  & \textbf{64.17} \\ \hline
GraphCodeBERT       & 85.16   & 62.85     \\
GraphCodeBERT\_Intr & 87.60 &62.26\\
{ContraBERT\_G} (MLM) &  87.30      &  62.01 \\
ContraBERT\_G (Contra)    &  85.63&  58.82  \\
{ContraBERT\_G}           & {90.06}  & {63.32} \\ \hline
\end{tabular}
\end{table}

\begin{table}[]
\centering
\caption{Results on code translation.}
\label{tbl-codetrans-coderefine}
\begin{tabular}{l|cc|cc}
\hline
 \multirow{3}{*}{Model} & \multicolumn{4}{c}{Code Translation} \\ \cline{2-5}
& \multicolumn{2}{c|}{Java $\rightarrow$ C\#} & \multicolumn{2}{c}{C\# $\rightarrow$ Java} \\
                      & BLEU-4        & Acc               & BLEU-4         & Acc  \\ \hline
% Na\"ive                  & 18.54        & 0                       & 18.69        & 0                       & 78.06   & 0       & 90.91   & 0          \\
% PBSMT                  & 43.53        & 12.50               & 40.06        & 16.10               &  -   &  -       &   -  &  -            \\
% LSTM                   & -            & -                      & -            & -                     & 76.76   & 10.00        & 72.08   & 2.50    \\
% Transformer            & 55.84            & 33.00                    & 50.47           & 37.90                  & 77.21   & 14.70    & 89.25   & 3.70    \\
% RoBERTa(code)          & 77.46        & 56.10              & 71.99        & 57.90               & -       & -         & -       & -         \\ \hline
CodeBERT  & 79.92        & 59.00                  & 72.14        & 58.00                 \\ 
CodeBERT\_Intr  & 79.93     & 59.20 & 75.71   &     58.60         \\ 
ContraBERT\_C (MLM)  &79.90&59.10  &75.03&   58.10   \\
ContraBERT\_C (Contra) & 51.99  &34.60   & 46.75  &38.30    \\
{ContraBERT\_C}             &    {79.95}    & {59.00}    &  {75.92}     &   {59.60}  \\ \hline
GraphCodeBERT    &80.58	  &59.40  &72.64	&58.80  \\
GraphCodeBERT\_Intr  & 80.61 &      59.60  &  75.50  & 60.10     \\ 
{ContraBERT\_G} (MLM) & 80.36 & {59.40}    & 75.10   &   {60.00}    \\
ContraBERT\_G (Contra) & 55.48 &39.40    & 48.92 &      39.00    \\
{ContraBERT\_G}  &    \textbf{80.78}  & \textbf{59.90}&  \textbf{76.24}  & \textbf{60.50} \\ \hline
\end{tabular}
\end{table}

\begin{table*}[]
\centering
\caption{Results on code search where the evaluation metric is MRR.}
\label{tbl-codesearch}
\begin{tabular}{lccccccc}
\hline
Model               & Ruby  & Javascript & Go    & Python & Java  & PHP   & Overall \\ \hline
% NBOW                & 0.162                     & 0.157                          & 0.330                     & 0.161                      & 0.171                     & 0.152  & 0.189                                          \\
% CNN                 & 0.276                     & 0.224                          &  0.680                    & 0.242                      & 0.263                     & 0.260  & 0.324                                          \\
% BiRNN               & 0.213                     & 0.193                          & 0.688                     & 0.290                      & 0.304                     & 0.338                     & 0.338                       \\
% SelfAtt             & 0.275                     & 0.287                          & 0.723                     & 0.398                      & 0.404                     & 0.426                     & 0.419                       \\
% RoBERTa             & 0.587                     & 0.517                          & 0.850                     & 0.587                      & 0.599                     & 0.560                     & 0.617                       \\
% RoBERTa(code)      & 0.628                     & 0.562                          & 0.859                     & 0.610                      & 0.620                     & 0.579                     & 0.643                       \\  \hline
CodeBERT            & 0.679                     & 0.620                          & 0.882                     & 0.672                      & 0.676                     & 0.628                     & 0.693                       \\  
CodeBERT\_Intr  &   0.686     &  0.623           &   0.883    &    0.676     &  0.678       &  0.630    &    0.696         \\ 
ContraBERT\_C (MLM) &0.675&0.621&0.888&0.670&0.675&0.631&0.693\\
ContraBERT\_C (Contra) &0.593&0.532&0.864&0.622&0.618&0.584&0.636\\
{ContraBERT\_C}          & {0.688}                     & {0.626}                        & {0.892}                     & {0.678}                      & {0.685}                     & {0.634}                     & {0.701}  \\      \hline    
GraphCodeBERT  & 0.703	&0.644	&0.897		&0.692	&0.691	&\textbf{0.649}	& 0.713 \\
GraphCodeBERT\_Intr  &  0.709     &   0.647 &  0.894     &   0.692 &    0.693  &0.647   & 0.714 \\ 
ContraBERT\_G (MLM) &0.692&0.642&0.897&0.690&0.690&0.647&0.710\\
ContraBERT\_G (Contra) &0.626&0.582&0.882&0.655&0.659&0.613&0.670\\
{ContraBERT\_G}          & \textbf{0.723}                     & \textbf{0.656}                        & \textbf{0.899}                     & \textbf{0.695}                      & \textbf{0.695}                     & {0.648}                     & \textbf{0.719}  \\      \hline    
\end{tabular}
\end{table*}

\subsection{RQ3: Performance of ContraBERT on Downstream Tasks.}\label{sec:rq-downstream}
We conduct extensive experiments on four downstream tasks to evaluate the performance of ContraBERT as compared to the original CodeBERT and GraphCodeBERT. We further add two baselines (i.e., CodeBERT\_Intr and GraphCodeBERT\_Intr), which are pre-trained by original data as well as the augmented variants. We supplement these two baselines to ensure the used scale of data is consistent with ContraBERT for a fair comparison. The results of clone/defect detection are shown in Table~\ref{tbl-clone-defect}. Table~\ref{tbl-codetrans-coderefine} presents the results of code translation and Table~\ref{tbl-codesearch} presents the results of code search where the rightmost ``overall'' column is the average value for six programming languages.  Because the values for clone detection and defect detection of GraphCodeBERT are not reported by their original paper~\cite{guo2020graphcodebert}, we use official code for reproduction and report these values. The other values of CodeBERT and GraphCodeBERT are directly taken from CodeXGLUE~\cite{lu2021codexglue} and Guo et al.~\cite{guo2020graphcodebert}.

From Table~\ref{tbl-clone-defect} and Table~\ref{tbl-codetrans-coderefine}, we find that ContraBERT\_C/G outperforms original pre-trained models CodeBERT or GraphCodeBERT on clone detection (POJ-104), defect detection and code translation. However, the absolute gains on code search (see Table~\ref{tbl-codesearch}) are minor. For these improvements, we attribute to the robustness-enhanced models providing better performance on downstream tasks. When it comes to minor improvements in code search, we ascribe to the difficulty of this task. Code search requires learning the semantic mapping between query and program. However, the semantic gap between programs and natural languages is huge. It makes the model difficult to achieve significant improvements. In total, considering the scale of testset on code search, which contains 52,561 samples for six programming languages, the improvements are still promising. Furthermore, we find that compared with CodeBERT and GraphCodeBERT, CodeBERT\_Intr and GraphCodeBERT\_Intr have better performance on these tasks. It is reasonable since we add extra data to further pre-train CodeBERT and GraphCodeBERT. However, the performance of CodeBERT\_Intr and GraphCodeBERT\_Intr is worse than ContraBERT\_C/G. It demonstrates that even with the same scale of data, ContraBERT\_C/G are still better than CodeBERT and GraphCodeBERT, which further strengthen our conclusion that the improvements are brought by our proposed approach rather than the gains brought by the increased scale of the data.

% ContraBERT outperforms CodeBERT by a significant margin in clone detection (from 82.67 to 90.46) and defect detection (from 62.08 to 64.17). Through the analysis of two tasks, we find that either clone detection or defect detection relies on learning program semantics. Especially for clone detection, it requires neural networks to learn the functionality of programs from different implementations. By designing the mutation operators and feeding the pair of similar data $(X, X')$ to the neural network, ContraBERT can learn a better representation space. 
% Table~\ref{tbl-codetrans-coderefine} shows the results of code translation, we can observe that ContraBERT\_C outperforms its original CodeBERT in BLEU-4 and Acc on both tasks. In addition, on code translation, ContraBERT\_G is still higher than GraphCodeBERT, but when it comes to code refinement, the performance is inconsistent. Specifically, BLEU-4 of ContraBERT\_G is lower than GraphCodeBERT, but Acc is higher than it. It is acceptable since different evaluation metrics tend to measure the quality of the generated sequence from different aspects (i.e., Acc measures the exact match with the ground-truth, BLEU-4 calculates the text similarity). Hence, it is hard for a model to obtain overwhelming results on all metrics. Generally, the overall performance on both tasks confirms ContraBERT is better than its original model.

\begin{tcolorbox}[breakable,width=\linewidth,
boxrule=0pt,top=1pt, bottom=1pt, left=1pt,right=1pt, colback=gray!20,colframe=gray!20]
\textbf{\ding{45} $\blacktriangleright$RQ3$\blacktriangleleft$} 
ContraBERT comprehensively improves the performance of original CodeBERT and GraphCodeBERT on four downstream tasks, we attribute the improvements to the enhanced robustness of the model has better performance on these tasks. 
\end{tcolorbox}

\subsection{RQ4: Ablation Study for Pre-training Tasks.}
ContraBERT utilizes two pre-training tasks, the first one is MLM, which learns the token representations and the second one is the contrastive pre-training task, which improves the model robustness by InfoNCE loss function. We further investigate the impact of each pre-training strategy on downstream tasks. The experimental results are shown in Table~\ref{tbl-clone-defect}, Table~\ref{tbl-codetrans-coderefine} and Table~\ref{tbl-codesearch} respectively, where the row of MLM or Contra denotes the results obtained by purely using MLM or contrastive pre-training task. For a fair comparison, the other settings are the same when combining both pre-training tasks for pre-training. 

We can observe that the performance of purely using contrastive pre-training tasks is worse than purely using MLM on these downstream tasks, especially on the task of code translation. It is acceptable since both pre-training tasks are excellent in different aspects. Specifically, MLM is designed by randomly masking some tokens in a sequence to help the model learn token representations. The learnt token representations are important for generation tasks to generate a target sequence such as code translation, so it will help the model achieve good performance on these tasks. However, the contrastive pre-training task is designed by grouping the semantically equivalent samples while pushing away the dissimilar samples through InfoNCE loss function. The model robustness is enhanced by the contrastive pre-training task. Furthermore, when combing both pre-training tasks, our model achieves better performance compared with purely using one of the pre-training tasks, which indicates that ContraBERT is robust at the same time is able to achieve better performance on the downstream tasks. 

\begin{tcolorbox}[breakable,width=\linewidth,
boxrule=0pt,top=1pt, bottom=1pt, left=1pt,right=1pt, colback=gray!20,colframe=gray!20]
\textbf{\ding{45} $\blacktriangleright$RQ4$\blacktriangleleft$} 
Masked language modeling (MLM) and the contrastive pre-training task play different roles for ContraBERT. When combining them together, the model achieves higher performance on different downstream tasks.
% Compared with InfoNCE,MLM is powerful to learn the token representations to benefit the downstream tasks, %however InfoNCE can also supplement it and promote MLM to shape a better representation space. Generally, ContraBERT achieves the best performance when combing MLM and InfoNCE objectives n the majority of the tasks. 
% however InfoNCE could also supplement it and reshape the vector space learnt by MLM. Hence, incorporating both objectives, ContraBERT achieves the best performance for the majority of the tasks. 
\end{tcolorbox}

\section{Discussion}
\label{sec-discussion}
In this section, we first discuss the implications of our work, then discuss the limitations followed by threats to validity.

\subsection{Implications}
In this work, we find that the widely used pre-trained code models such as CodeBERT~\cite{feng2020codebert} or GraphCodeBERT~\cite{guo2020graphcodebert} are not robust to adversarial attacks. Based on this finding, we further propose a contrastive learning-based approach for improvement. We believe that this finding in our paper will inspire the following-up researchers when designing a new model architecture for code, considering some other problems in the model such as robustness, generalization and not just focusing on the accuracy of the model on different tasks.

% \subsection{Threats to Validity}
\subsection{Limitations}
By our experimental results, we find that the robustness of the model is enhanced significantly compared with the original models. We attribute it to the contrastive pre-training task to learn the semantically equivalent samples. However, these robustness-enhanced models only have slight improvements on the downstream task of code search. For this task, since it requires learning the semantic mapping between a query and its corresponding code, the designed augmentation operators just modify the code or query itself, hence their correlations are not captured and this leads to the improvements being limited. For code search, a possible solution to further improve the performance is to build the token relations between PL and NL for augmented variants, however, it involves intensive work to analyse the relations between the program and natural language comment. We will explore it in our future work. 

Another limitation is the designed augmentation operators for PL and NL. We just design some basic operators to transform programs and comments. These operators are straightforward, although they are confirmed their effectiveness in improving model robustness. It is intriguing to explore more complex augmentation strategies such as multiple operations on these operators for a sample to construct complex augmented variants.

\subsection{Threats to Validity}
\noindent \textbf{Internal validity:} The first threat is the hyper-parameter tuning for pre-training. More hyper-parameters need to tune than CodeBERT or GraphCodeBERT for example the temperature $t$, the momentum coefficient $m$ and \textit{queue} size. We follow the original settings from MoCo~\cite{he2020momentum} and these parameters may not be optimal as they are designed for the task of image classification in computer vision. Due to that, the pre-training process is time-consuming and resource-consuming. We need nearly 2 days to complete one training process hence we ignore the hyper-parameter tuning process. However, we also find that even with the original parameters used in MoCo~\cite{he2020momentum}, ContraBERT still achieves higher performance than the original models. The second threat is that we use the same train-validation-test split that CodeXGLUE~\cite{lu2021codexglue} and GraphCodeBERT~\cite{guo2020graphcodebert} used. Adjusting the data split ratio or improving the training data quality may produce better results, however, we do not take these strategies to ensure a fair evaluation. The third threat is that we just use clone detection(POJ-104) to verify the robustness of the model is enhanced in Fig.~\ref{fig:robust} and Section~\ref{sec:rq1}, we also plot the learnt space in Section~\ref{sec:rq-visualization}. The reason to select clone detection is that it aims at identifying the semantically equivalent programs from other distractors, which is suitable for the evaluation.

\noindent \textbf{External validity:} Some other pre-training works in the code scenario such as CuBERT~\cite{kanade2019pre} are not included for evaluation. CuBERT was pre-trained on a large Python corpus with MLM. Our approach is orthogonal to these pre-trained models and we just need to replace the encoder M with other existing pre-trained models for evaluation.

\section{Related Work}
\label{sec-related}
In this section, we briefly introduce the related works on contrastive learning, the pre-trained models for ``big code'' and the adversarial robustness of models of code.
\subsection{Contrastive Learning}
Contrastive learning is to learn representations by minimizing the distance between similar samples while maximizing the distance between different samples to help the similar samples closer to each other and different samples far apart from each other. Over the past few years, it has attracted increasing attention with many successful applications in computer vision~\cite{he2020momentum, chen2020simple, kim2020adversarial, radford2021learning, dai2017contrastive}, natural language processing~\cite{yang2019reducing, fang2020cert, gao2021simcse, shen2020simple}. Recently, there are some works~\cite{bui2021self, chen2021varclr, jain2020contrastive} that utilize contrastive learning for different software engineering tasks. For example, Bui et al.~\cite{bui2021self} proposed Corder, a contrastive learning approach for code-to-code retrieval, text-to-code retrieval and code-to-text summarization. VarCLR~\cite{chen2021varclr} aimed to learn the semantic representations of variable names based on contrastive learning for different downstream tasks such as variable similarity scoring and variable spelling error correction. ContraCode~\cite{jain2020contrastive} generated variants by a source-to-source compiler on JavaScript and further combined these generated mutated samples with contrastive learning for the task of clone detection, type inference and code summarization. Compared with these existing works which only focus on designing mutated variants for code, we first illustrate the widely concerned CodeBERT and GraphCodeBERT are weak to the adversarial examples. Then we design a set of simple and complex augmented operators on both programs and natural language sequences to obtain different variants. By contrastive learning to learn semantically equivalent variants, the robustness of existing pre-trained models is enhanced. We further confirm that the robustness-enhanced models provide improvements on different downstream tasks.

\subsection{Pre-trained Models for ``Big Code''}
Recently, pre-trained models are widely applied to the ``big code'' era~\cite{kanade2019pre, feng2020codebert, guo2020graphcodebert, lu2021codexglue, svyatkovskiy2020intellicode,buratti2020exploring, karampatsis2020scelmo, wang2021codet5, ahmad2021unified, jain2020contrastive, liu2022commitbart}. For example, Kanade et al.~\cite{kanade2019pre} pre-trained CuBERT based on BERT~\cite{devlin2018bert} with a massive corpus of Python programs from GitHub and then fine-tuned it for some classification tasks such as variable misuse classification. Feng et al.~\cite{feng2020codebert} proposed CodeBERT, a bimodal pre-trained model for programming language (PL) and natural language (NL) that learns the program representation to support code search and source code summarization. GraphcodeBERT~\cite{guo2020graphcodebert} combines the variable data-flow graph in a program with the code sequences and the natural language sequence to enhance CodeBERT. CodeXGLUE~\cite{lu2021codexglue} also utilized CodeBERT and CodeGPT~\cite{radford2019language} to release a benchmark including several software engineering tasks. Liu et al.~\cite{liu2022commitbart} proposed a CommitBART to support commit-related downstream tasks. Compared with existing pre-trained models, we illustrate they are not robust and further propose ContraBERT to enhance model robustness. 

\subsection{Adversarial Robustness on Models of Code}
The research about adversarial robustness analysis on the models of code has attracted the attention~\cite{srikant2021generating, yefet2020adversarial, ramakrishnan2020semantic, yang2022natural, bielik2020adversarial, zhang2020generating}. Generally, these works can be categorized into two groups: white-box and black-box manner, where the white-box means that the approach provides some explanations on the decision-making  while the black-box mainly focuses on the statistical evaluation. In terms of white-box works, Yefet et al.~\cite{yefet2020adversarial} proposed DAMP to select the semantic preserving perturbations by deriving the output distribution of the model with the input. Srikant et al.~\cite{srikant2021generating} provided a general formulation of a perturbed program that models site locations and perturbation choices for each location. Then based on this formulation, they further proposed a set of first-order optimization algorithms for the solving. In terms of the black-box works, HMH~\cite{zhang2020generating} generated adversarial examples of the source code by conducting iterative identifier renaming and evaluated on source code functionality classification task. The latest work by Yang et al.~\cite{yang2022natural} proposed ALERT to transform the inputs while preserving the optional semantics of original inputs by replacing the variables with the substitutes. Their experiments are conducted on the pre-trained models CodeBERT and GraphCodeBERT. Compared with ALERT, which only designed the rename variable operation, in this paper, apart from the rename variable operation, we further design eight augmented operators over PL-NL pairs. Furthermore, a newly designed model to solve the weakness of robustness is not involved in ALERT. In contrast, we propose our general network architecture that uses contrastive learning to enhance model robustness. The extensive experiments have confirmed that our approach enhances the robustness of existing pre-trained models. We also demonstrate that these robustness-enhanced models provide improvements on different downstream tasks.

\section{Conclusion}
\label{sec-con}
In this paper, we observe that state-of-the-art pre-trained models such as CodeBERT and GraphCodeBERT are not robust to adversarial attacks and a simple mutation operator (e.g., variable renaming) degrades their performance significantly. To address this problem, in this paper, we propose ContraBERT, a contrastive learning-based framework to enhance the robustness of existing pre-trained models by designing nine kinds of PL-NL augmented operators to group the semantically equivalent variants. Through extensive experiments, we have confirmed that the model's robustness is enhanced. Furthermore, we also illustrate that these robustness-enhanced models provide improvements on four downstream tasks.

\section{ACKNOWLEDGMENTS}
We express our sincere gratitude to Mr Daya Guo from Sun Yat-sen University for his assistance. This research is partially supported by the National Research Foundation, Singapore under its the AI Singapore Programme (AISG2-RP-2020-019), the National Research Foundation, Prime Ministers Office, Singapore under its National Cybersecurity R\&D Program (Award No. NRF2018NCR-NCR005-0001), NRF Investigatorship NRF-NRFI06-2020-0001, the National Research Foundation through its National Satellite of Excellence in Trustworthy Software Systems (NSOE-TSS) project under the National Cybersecurity R\&D (NCR) Grant award no. NRF2018NCR-NSOE003-0001, the Ministry of Education, Singapore under its Academic Research Tier 3 (MOET32020-0004). IIE authors are supported in part by NSFC (61902395), Beijing Nova Program. Any opinions, findings and conclusions or recommendations expressed in this material are those of the author(s) and do not reflect the views of the Ministry of Education, Singapore.

\bibliographystyle{IEEEtran}
\bibliography{ref}
\end{document}